\documentclass[11pt]{article}
\usepackage[greek,francais, english]{babel}
\usepackage[round]{natbib}
\usepackage{a4}
\usepackage{color}
\usepackage{hyperref}
\usepackage{graphicx}
\usepackage{array, blindtext}
\usepackage{caption}
\usepackage{lscape}
\extrafloats{100}
\usepackage{tablefootnote}
\usepackage{multirow}
\def\GEOSauth#1 {\uppercase{ #1} \vskip 0.3cm plus 0.1cm minus 0.1cm}
\newcounter{authno}
\def\GEOSinst#1 {\small \addtocounter{authno}{1} $^{\arabic{authno}}$ #1
        \vskip 1mm \large}
{\bf \def\GEOSx#1 {\small \ts $^*$ #1 \vskip 1mm \large}
\def\GEOSinsto#1 {\small \ts #1 \vskip 1mm \large}
\title{\vspace{-2.5cm}
\large
{\bf GEOS RR 62  \hspace{2cm} GEOS CIRCULAR ON RR LYRAE  \hspace*{\fill}  {June 15, 2023}} \\
\vspace{1.5cm}
\LARGE
\LARGE
{\uppercase {\bf Observations of the suspected RR Lyr stars NSV 14172 and NSV 14264}}
\vspace{0.5cm}
}
\author{J.F. Le Borgne$^{1,2,3}$
}
\date{\vspace{-5ex}}    % remove the date in title using \maketitle
\begin{document}
\maketitle
\GEOSinst{GEOS (Groupe Europ\'een d'Observations Stellaires), http://geos.upv.es}
\GEOSinst{IRAP; OMP; Universit\'e de Toulouse; 14, avenue Edouard Belin, F-31400 Toulouse, France}
\GEOSinst{LAM, Laboratoire d'Astrophysique de Marseille, 38 Rue Frédéric Joliot Curie, F-13013 Marseille, France}
\vspace{5mm}
\begin{abstract}
NSV 14264 and NSV 14172 are suspected to be variable stars of RR Lyr type \citep{Brun}.
They were observed during three nights in October 2018 with a 25cm diameter
telescope. These observations completed by ASAS-SN survey data bring to the
conclusion that these two stars are not RR Lyraes but constant stars in
the limit of the precision of the present photometry. The analysis of GAIA data allows
to say that NSV 14264 is a main sequence dwarf similar to the Sun but that NSV 14172 is
a yellow giant star located in the HR diagram at the limit between RR Lyraes and CW cepheids;
however, it does not pulsate with significant amplitude.
\end{abstract}
\section{Introduction}
NSV 14172 and NSV 14264 were introduced in the suspected variable catalog
\citep{GCVS} after the publication by A. Brun of a list of variable star
candidates \citep{Brun}. These 2 stars were suspected to be of RR Lyr type
and have numbers 49 and 59 respectively in \cite{Brun}. The observations were done
by R. Weber between August 1959 and December 1962 using a photographic camera.
The range of photographic magnitudes are 12.5 to 13.6 and 12.2 to 13.4 respectively.
It is surprising that such bright stars are still suspected and that no publication
reporting on their real status is available. In the present paper we report
observations of these two stars made on October 3, 4 and 5, 2018 in order to
clarify their photometric properties.
\section{Observations}
The observations were made in J.F. Le Borgne's private observatory (EsO) in
Escalquens (Occitania, EU) using a 10 inches diameter newton f/4 telescope
(Skywatcher) equipped with a CCD camera (Apogee Alta F9000, KAF-09000) and
optical aberration corrector giving a field of 2$^{\circ}\times$2$^{\circ}$.
A Johnson R filter was used with an exposure time of one minute.\\
Usual dark and flat-field corrections were done with the use of the software {\it IRAF}
running on a fedora linux system. Astrometry was performed using {\it imwcs} part
of {\it WCSTools} package\footnote{\href{http://tdc-www.harvard.edu/wcstools/}
{http://tdc-www.harvard.edu/wcstools/}}
and photometry using {\it SExtractor} program \citep{Bertin}.\\
NSV 14264 was observed on October 3 and 4, 2018 and NSV 14172 on October 5, 2018
(table \ref{tab1}). Each night, images were continuously acquired during more than
10 hours. The number of images obtained are between 439 and 457 per night.
The comparison and check stars for each star are given in table \ref{tab2} were
data from UCAC4 catalog \citep{Zacharias} are given. The magnitudes of the
studied stars are obtained by adding the UCAC4 SDSS r magnitude of comparison
star to instrumental magnitude differences.
\begin{table}[htbp]
\begin{center}
\small
\begin{tabular}{lllllll}
\hline
          &         &          &           &           &           &  \\
Date      & JD      & duration & star      & Mean      & standard  & Number of \\
2018      &         &          &           & magnitude & deviation & measurements \\
          &         &          &           &           &           &  \\
\hline
          &         &          &           &           &           &  \\
October 3 & 2458395 & 10h10mn  & NSV 14264 & 11.459    & 0.011     & 439 \\
October 4 & 2458396 & 10h32mn  & NSV 14264 & 11.460    & 0.010     & 457 \\
October 5 & 2458397 & 10h03mn  & NSV 14172 & 12.425    & 0.013     & 445 \\
          &         &          &           &           &           &  \\
\hline
\end{tabular}
\end{center}
\vspace{-5mm}
\caption{\label{tab1} Nightly information about observations.}
\end{table}
\normalsize
\begin{table}[htbp]
\begin{center}
\small
\begin{tabular}{llcccccc}
\hline
           &            &               &               &        &        &        &        \\
           &   UCAC4    &  ra(J2000)    &   dec(J2000)  &  B     &  V     & r      & B-V    \\
           &            &               &               &        &        &        &        \\
\hline
           &            &               &               &        &        &        &        \\
 NSV 14172 & 674-114332 & 22:28:42.9422 & +44:39:42.240 & 13.411 & 12.692 & 12.477 & 0.719  \\
 comp. star& 674-114331 & 22:28:42.6767 & +44:45:36.732 & 13.252 & 12.678 & 12.527 & 0.574  \\
 check star& 674-114345 & 22:28:50.5750 & +44:38:58.040 & 13.413 & 12.857 & 12.700 & 0.556  \\
           &            &               &               &        &        &        &        \\
\hline
           &            &               &               &        &        &        &        \\
 NSV 14264 & 685-122818 & 22:39:04.7963 & +46:49:57.021 & 12.949 & 11.983 & 11.700 & 0.966  \\
 comp. star& 685-122869 & 22:39:20.9711 & +46:48:07.358 & 11.464 & 11.362 & 11.473 & 0.111  \\
 check star& 685-122744 & 22:38:43.9073 & +46:54:55.862 & 11.759 & 11.458 & 11.461 & 0.003  \\
           &            &               &               &        &        &        &        \\
\hline
\end{tabular}
\end{center}
\vspace{-5mm}
\caption{\label{tab2} Coordinates (equinox J2000.0) and UCAC4 photometry of NSV 14172,
NSV 14264 and their comparison and check stars.}
\end{table}
\normalsize
\section{Light curves}
\begin{figure}
\begin{center}
\begin{tabular}{p{8cm}p{8cm}}
\vspace{-9mm}\begin{center} \includegraphics[width=7cm]{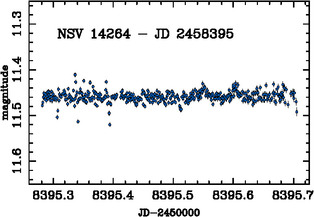} \end{center}&
\vspace{-9mm}\begin{center} \includegraphics[width=7cm]{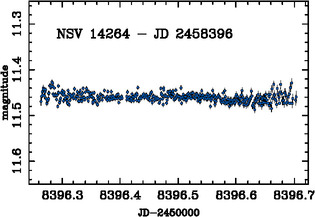} \end{center}\\
\vspace{-9mm}\begin{center} \includegraphics[width=7cm]{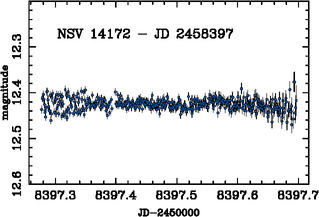} \end{center}&
\vspace{1mm} \captionof{figure}{\label{lcOct2018} R light curves of NSV 14264
and NSV 14172 on October 3, 4 and 5, 2018. Time scales are heliocentric julian
days.} \\
\end{tabular}
\end{center}
\end{figure}

In figure \ref{lcOct2018} the light curves of the two stars are drawn for the three nights.
All three graphs have a full scale of 0.4 magnitudes in ordinate. The light curves show no
significant variation for the three nights and the two stars. Since all time series cover
about 10 hours (duration of the order of magnitude of the period of RR Lyr stars), it can be
said that these stars cannot be RR Lyraes. Table \ref{tab1} gives the mean magnitudes per
night and standard deviations. The standard deviations are of the order of 0.01 magnitude.
Table \ref{tab3} gives the mean magnitude and standard deviations for the whole run. \\

\begin{figure}
\begin{center}
\begin{tabular}{cc}
\includegraphics[width=7cm]{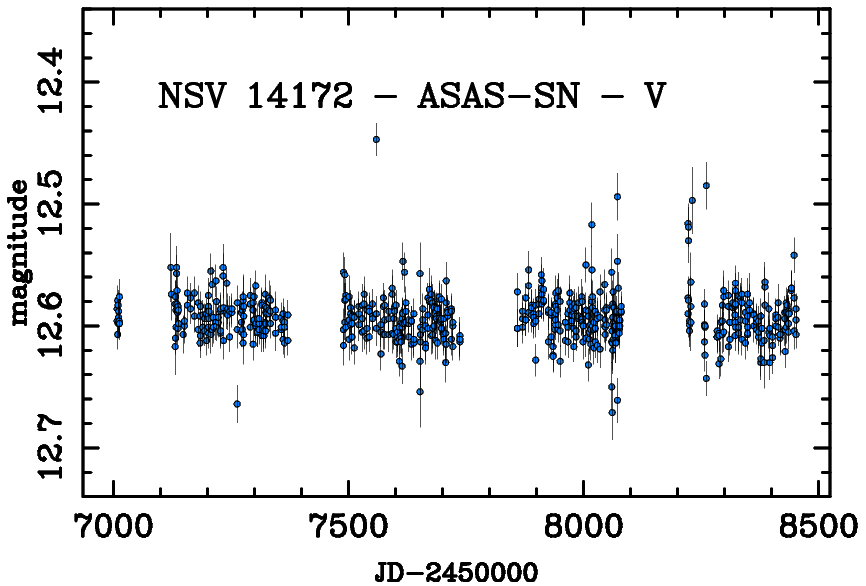}&
\includegraphics[width=7cm]{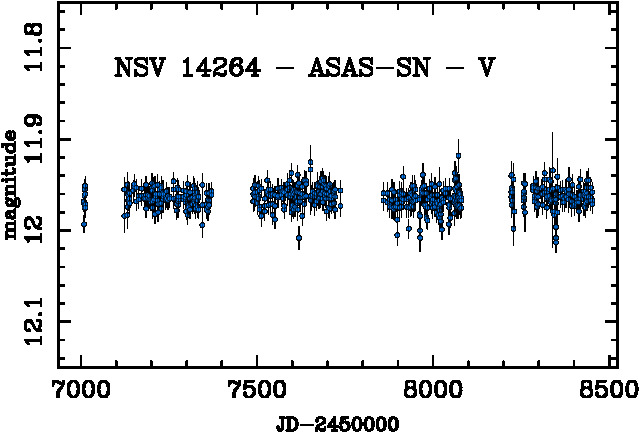}\\
\includegraphics[width=7cm]{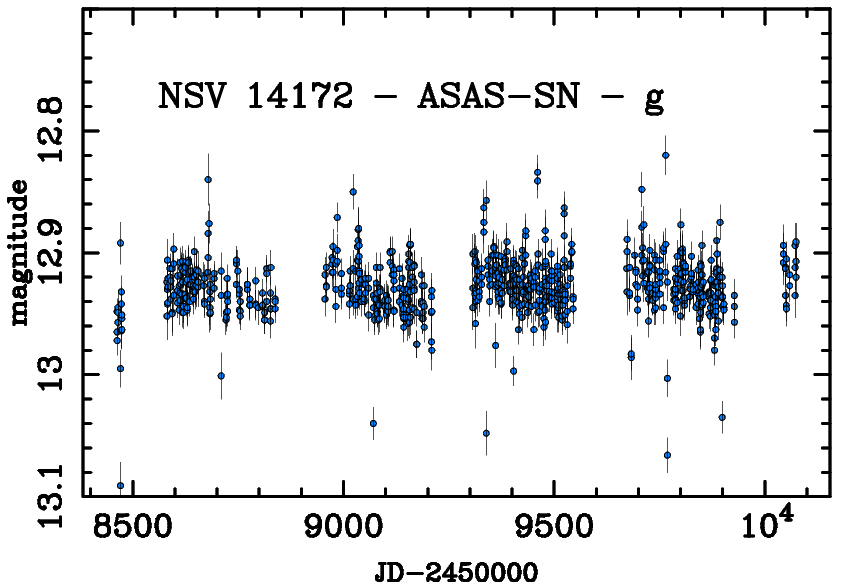}&
\includegraphics[width=7cm]{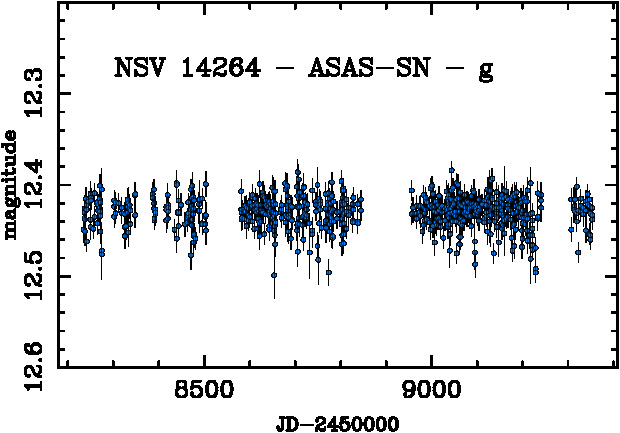}\\
\end{tabular}
\end{center}
\vspace{-5mm}
\caption{\label{lcASASSN} V and g light curves of NSV 14172 and NSV 14264 from
ASAS-SN. Time scales are heliocentric julian days.}
\end{figure}

\begin{table}[htbp]
\begin{center}
\small
\begin{tabular}{llllcc}
\hline
           &            &               &                &           &            \\
           &   UCAC4    & filter        & number of meas.& Mean      & Standard   \\
           &            &(observatory)  & (of nights)    & magnitude & deviation  \\
           &            &               &                &           &            \\
\hline
           &            &               &                &           &            \\
 NSV 14172 & 674-114332 & R (EsO)       & 445 (1)        & 12.425    & 0.013      \\
 check star& 674-114345 & R (EsO)       & 445 (1)        & 12.703    & 0.014      \\
           &            &               &                &           &            \\
 NSV 14172 & 674-114332 & V (ASAS-SN)   & 627 (226)      & 12.593    & 0.019      \\
 NSV 14172 & 674-114332 & g (ASAS-SN)   & 792 (252)      & 12.928    & 0.026      \\
           &            &               &                &           &            \\
\hline
           &            &               &                &           &            \\
 NSV 14264 & 685-122818 & R (EsO)       & 896 (2)        & 11.459    & 0.011      \\
 check star& 685-122744 & R (EsO)       & 896 (2)        & 11.407    & 0.013      \\
           &            &               &                &           &            \\
 NSV 14264 & 685-122818 & V (ASAS-SN)   & 628 (226)      & 11.966    & 0.013      \\
 NSV 14264 & 685-122818 & g (ASAS-SN)   & 814 (260 )     & 12.430    & 0.015      \\
           &            &               &                &           &            \\
\hline
\end{tabular}
\end{center}
\vspace{-5mm}
\caption{\label{tab3} Mean magnitudes of NSV 14172, NSV 14264 and their check
stars.}
\end{table}
\normalsize
In order to confirm these findings, we analyzed the data from the survey ASAS-SN \citep{Shappee} 
of both stars. This survey takes only a couple of measurements per night but the data cover several 
years. Two filters are available, V and g. V filter (2014-2018) and g filter (2018-2023) have only 
7 months of in common from April to November 2018. The light curves in V and g are plotted in figure
\ref{lcASASSN} (as in figure \ref{lcOct2018}, the ordinate full scale is 0.4 magnitudes). The mean 
values and standard deviations are shown in table \ref{tab3}. The ASAS-SN data confirm that these
stars have no significant light variation within a range of a couple of hundredths of magnitude 
($3\sigma\sim0.07$ mag.) for each filter and over the time interval of four years. The data from 
ASAS-SN used for the present statistics were cleaned for internal error greater than 0.02 magnitudes 
in g and 0.03 in V for NSV 14172,  0.03 in g and V for NSV 14264. Only data from camera "bC" were used 
for NSV 14172 and NSV 14264 for filter g. Data from camera "bs", though contemporary to camera "bC", 
were more dispersed. The camera used for both stars in V is "bc" only.
\section{Characterizing UCAC4 674-114332 and UCAC4 685-122818}
Table \ref{tab4} summarizes the values of physical parameters of NSV 14172, NSV 14264 as deduced
from GAIA observations, DR2 and DR3, published by the \cite{GaiaDR2,GaiaDR3} and consequent papers.
It appears that the two stars are very different. NSV 14172 is a dwarf quite similar to the Sun
with an effective temperature of 5400 -- 5900 K, a stellar radius of about 1.1 R$_{\odot}$ and
a mass of 1 M$_{\odot}$, slightly less luminous than the Sun (0.8-0.9 L$_{\odot}$).
NSV 14264, on the other hand, is a yellow giant star 24 times more luminous than the Sun and a radius
of 7.38 or 7.818 R$_{\odot}$. Its effective temperature is about 4900 K. This difference of class,
since their magnitudes are quite close to each other, is reflected in their distances: 320 pc for
NSV 14172, 1228-1274 pc for NSV 14264, in accordance with the measured parallaxes of 3.1063 mas and
0.7504 mas from Gaia DR3. The estimated gravities ${log(g)}$ are also characteristic of their respective
classes. \\
The G absolute magnitudes published by \cite{Anders2022} are 4.48 for NSV 14172 (\cite{GaiaDR3} gives
4.75) and 1.02 for NSV 14264.
These values with a BP-RP value of 0.9427 place NSV 14172 on the main sequence as expected while the
BP-RP value of 1.2217 for NSV 14264 places it among the RR Lyraes as can be seen in figure 3 of
\cite{Eyer}. The published effective temperature is cooler than typical temperatures of RR Lyraes, though,
more typical of cepheids. However, NSV 14264 does not pulsate with expected amplitude. 
\begin{table}[htbp]
\begin{center}
\small
\begin{tabular}{llll}
\hline
               &                       &                           &                          \\
{\bf Parameter}&  {\bf NSV 14172}      &  {\bf NSV 14264}          &  {\bf Source}            \\
               &                       &                           &                          \\
plx (mas)      &  3.1063$\pm$0.0132    &  0.7504$\pm$0.0128        &  \cite{GaiaDR3}          \\
               &                       &                           &                          \\
dist. (pc)     &  310.577 (307.7-313.5)&  1228.878  &  \cite{Bailer-Jones2018} \\
               &  310.5780$\pm$2.883   &  1274.9399$\pm$47.820     &  \cite{TESS2019}         \\
               &  320.136 (318.6-321.5)&  1271.854 &  \cite{Bailer-Jones2021} \\
               &  320.7386             &                           &  \cite{GaiaDR3}          \\
               &  320.462              &  1283.0                   &  \cite {Anders2022}      \\
               &                       &                           &                          \\
T$_{eff}$ (K)  &  5374.33              &  4937.93                  &  \cite{GaiaDR2}          \\
               &  5377.6               &                           &  \cite{GaiaDR3}          \\
               &  5735$\pm$154         &  5155$\pm$175             &  \cite{Bai2019}          \\
               &  5907.41              &  4843.02                  &  \cite {Anders2022}      \\
               &                       &                           &                          \\
log(g) (cm/s2) &  4.3143               &                           &  \cite{GaiaDR3}          \\
               &  4.4087               &  2.7795                   &  \cite {Anders2022}      \\
               &                       &                           &                          \\
 $[Fe/H]$      &  -0.7598              &                           &  \cite{GaiaDR3}          \\
               &  -0.0396              &  -0.1154                  &  \cite {Anders2022}      \\
               &                       &                           &                          \\
Radius (R$_{\odot}$)&  1.05            &  7.38                     &  \cite{GaiaDR2}          \\
               &  1.1332               &                           &  \cite{GaiaDR3}          \\
               &  1.018                &  7.818                    &  \cite{TESS2019}         \\
               &                       &                           &                          \\
Luminosity (L$_{\odot}$)&  0.826       &  29.159                   &  \cite{GaiaDR2}          \\
               &  0.9659               &                           &  \cite{GaiaDR3}          \\
               &                       &                           &                          \\
Mass (M$_{\odot}$)& 0.901              &                           &  \cite{GaiaDR3}          \\
               &  0.996                &                           &  \cite{TESS2019}         \\
               &  0.999                &  1.376                    &  \cite {Anders2022}      \\
               &                       &                           &                          \\
class          &  dwarf                &  giant                    &  \cite{TESS2019}         \\
               &                       &                           &                          \\
G Abs. mag.    &  4.7556               &                           &  \cite{GaiaDR3}          \\
               &  4.4833               &  1.0264                   &  \cite {Anders2022}      \\
               &                       &                           &                          \\
mag G          &  12.3951$\pm$0.0002   &  11.7418$\pm$0.0001       &  \cite{GaiaDR3}          \\
mag BP         &  12.7822$\pm$0.0005   &  12.2755$\pm$0.0004       &  \multicolumn{1}{c}{"}   \\
mag RP         &  11.8394$\pm$0.0003   &  11.0539$\pm$0.0003       &  \multicolumn{1}{c}{"}   \\
BP-RP          &  0.9427               &  1.2217                   &  \multicolumn{1}{c}{"}   \\
               &                       &                           &                          \\
\hline
\end{tabular}
\end{center}
\vspace{-5mm}
\caption{\label{tab4} Physical parameters and photometry (rounded to 0.0001 mag) of NSV 14172 and
NSV 14264 deduced from GAIA data.}
\end{table}
\normalsize

UCAC4 685-122818 (NSV 14724) was investigated by \cite{Cantat-Gaudin2018} as possibly
belonging to the open cluster ASCC 124 also known as Alessi 37 \citep{Kharchenko}
They estimated the probability of stars to belong to clusters from GAIA DR1 data. The
result was that UCAC4 685-122818 does not belong to ASCC 124 / Alessi 37.
\section{Checking for misidentification}
It is quite common that variable stars are misidentified in discovery papers, giving wrong
coordinates. \cite{Brun} provides finding charts: we have compared these charts with the
Digital Sky Survey (DSS) as provided by the European Southern Observatory
\footnote{\href{http://archive.eso.org/dss/dss/}http://archive.eso.org/dss/dss/}.
We choose to make the comparison with the blue DSS images expecting a better similarity of wavelength
response of photographic plates. The comparison is not straightforward, but we
can say that the identifications of Brun-49 with UCAC4 674-114332 and Brun-59
with UCAC4 685-122818 is good with a reasonable confidence.\\
Since the misidentification could have been done on the drawing of the finding
charts, we have examined the light curves of the stars in the field obtained
at EsO. We have drawn the light curves of about a thousand stars in the
fields of  UCAC4 674-114332 and of UCAC4 685-122818.
None of them shows light variations characteristic of RR Lyraes.
The selection criteria was that they were brighter than 15 and fainter than 12 in V filter, and
with B-V less than 0.9 in UCAC4 catalog \citep{Zacharias}.
\section{Conclusion}
We have examined the possible light variation of two RR Lyrae candidates included in the list of 
\cite{Brun}. 
None of them were proved to vary as noticed from our own observations and from ASAS-SN archive data.
From GAIA data we can say that NSV 14172 is a bona fide one solar mass main sequence star while 
NSV 14264 is a yellow giant. However, NSV 14264 luminosity and effective temperature are typical of 
a star located in the pulsating star instability strip.
\section{Acknowledgements}
The present study makes use of the following facilities: \\
- The GEOS database of RR Lyr stars hosted at Institut de Recherche en Astrophysique et
Planétologie, Toulouse, France http://rr-lyr.irap.omp.eu/dbrr/ \\
- The SIMBAD database, operated at CDS, Strasbourg,France \citep{Wenger}, of the VizieR
catalogue access tool also at CDS. The original description of the VizieR service was published
in A\&A, Supp. 143, 23, http://vizier.u-strasbg.fr/viz-bin/VizieR and of "Aladin sky atlas" 
developed at CDS, Strasbourg Observatory, France \citep{Bonnarel,Boch}\\
- The International Variable Star Index (VSX)
database, operated at AAVSO, Cambridge, Massachusetts, USA\\
- Data products from the AAVSO Photometric All Sky Survey (APASS). Funded by the Robert Martin Ayers
Sciences Fund and the National Science Foundation.


\begin{thebibliography}{}
\bibitem[Anders et al., 2019]{Anders2019}Anders, F., Khalatyan, A., Chiappini, et al.\ 2019, A\&A, 628A, 94
\href{https://ui.adsabs.harvard.edu/abs/2019A\%26A...628A..94A}{[NASA-ADS]}
\bibitem[Anders et al., 2022]{Anders2022}Anders, F., Khalatyan, A., Queiroz, A.B.A., et al.\ 2022, A\&A, 658, A91
\href{https://ui.adsabs.harvard.edu/abs/2022A\%26A...658A..91A}{[NASA-ADS]}
\bibitem[Bai et al., 2019]{Bai2019}Bai Y., Liu J., Bai Z., Wang S., Fan D.\ 2019, AJ, 158, 93
\href{https://ui.adsabs.harvard.edu/abs/2019AJ....158...93B}{[NASA-ADS]}
\bibitem[Bailer-Jones et al., 2018]{Bailer-Jones2018}Bailer-Jones, C.A.L., Rybizki, J., Fouesneau, M., Mantelet, G., Andrae R.\ 2018, AJ, 156, 58
\href{https://ui.adsabs.harvard.edu/abs/2018AJ....156...58B}{[NASA-ADS]}
\bibitem[Bailer-Jones et al., 2021]{Bailer-Jones2021}Bailer-Jones, C.A.L., Rybizki, J., Fouesneau, M., Demleitner, M., Andrae R.\ 2021, AJ, 161, 147
\href{https://ui.adsabs.harvard.edu/abs/2021AJ....161..147B}{[NASA-ADS]}
\bibitem[Bertin \& Arnouts, 1996]{Bertin} Bertin, E., Arnouts, S.\ 1996, A\&A, 117, 393
\href{https://ui.adsabs.harvard.edu/abs/1996A\%26AS..117..393B}{[NASA-ADS]}
\bibitem[Boch \& Fernique, 2014]{Boch}Boch, T., Fernique, P.\ 2014, ASP Conf. Ser., 485, 277
\href{https://ui.adsabs.harvard.edu/abs/2014ASPC..485..277B}{[NASA-ADS]}
\bibitem[Bonnarel et al., 2000]{Bonnarel}Bonnarel, F., Fernique, P., Bienaymé, O., et al.\ 2000, A\&AS, 143, 33
\href{https://ui.adsabs.harvard.edu/abs/2000A\%26AS..143...33B}{[NASA-ADS]}
\bibitem[Brun, 1964]{Brun}Brun, A.\ 1964, Journal des Observateurs, 47, 45
\href{https://ui.adsabs.harvard.edu/abs/1964JO.....47...45B}{[NASA-ADS]}
\bibitem[Cantat-Gaudin et al., 2018]{Cantat-Gaudin2018}Cantat-Gaudin, T., Vallenari, A., Sordo, R., et al.\ 2018, A\&A, 615A, 49
\href{https://ui.adsabs.harvard.edu/abs/2018A\%26A...615A..49C}{[NASA-ADS]}
\bibitem[Gaia Coll., 2018]{GaiaDR2}Gaia Collaboration, Brown A.G.A., Vallenari A., Prusti T., et al.\ 2018, A\&A, 616, 1
\href{https://ui.adsabs.harvard.edu/abs/2018A\%26A...616A...1G}{[NASA-ADS]}
\bibitem[Gaia Coll., 2019]{Eyer}Gaia Collaboration, Eyer, L., Rimoldini, L., Audard, M., Anderson, R. I., Nienartowicz, K.\ 2019, A\&A, 623A, 110
\href{https://ui.adsabs.harvard.edu/abs/2019A\%26A...623A.110G/abstract}{[NASA-ADS]}
\bibitem[Gaia Coll., 2022]{GaiaDR3}Gaia Collaboration\ 2022, yCat.1355....0G %(DR3)
\href{https://ui.adsabs.harvard.edu/abs/2022yCat.1355....0G}{[NASA-ADS]}
\bibitem[Kharchenko et al., 2005]{Kharchenko} Kharchenko, N. V., Piskunov, A. E., R\"oser, S., Schilbach, E., Scholz R.-D.\ 2005, A\&A 440, 403
\href{https://ui.adsabs.harvard.edu/abs/2005A\%26A...440..403K}{[NASA-ADS]}
\bibitem[Samus et al., 2017]{GCVS}Samus N.N., Kazarovets E.V., Durlevich O.V., Kireeva N.N., Pastukhova E.N.,
General Catalogue of Variable Stars: Version GCVS 5.1\ 2017, Astronomy Reports 61, No. 1, 80
\href{https://ui.adsabs.harvard.edu/abs/2017ARep...61...80S}{[NASA-ADS]}
\bibitem[Shappee et al., 2014]{Shappee}Shappee, B.J., Prieto, J.L., Grupe, D., et al.\ 2014, ApJ, 788, 48 % ASAS-SN
\href{https://ui.adsabs.harvard.edu/abs/2014ApJ...788...48S}{[NASA-ADS]}
\bibitem[Stassun et al., 2019]{TESS2019}Stassun K.G., Oelkers R.J., Paegert M., et al.\ 2019, AJ, 158, 138
\href{https://ui.adsabs.harvard.edu/abs/2019AJ....158..138S}{[NASA-ADS]}
\bibitem[Wenger et al., 2000]{Wenger}Wenger, M., Ochsenbein, F., Egret, D., et al.\ 2000, A\&A Supp., 143, 9
\href{https://ui.adsabs.harvard.edu/abs/2000A\%26AS..143....9W}{[NASA-ADS]}
\bibitem[Zacharias et al., 2013]{Zacharias}Zacharias N., Finch C. T., Girard T. M., et al.\ 2013, AJ, 145, 44
\href{https://ui.adsabs.harvard.edu/abs/2013AJ....145...44Z}{[NASA-ADS]}
\end{thebibliography}
\end{document}